%% file: main.tex
\documentclass[conference]{IEEEtran}
\IEEEoverridecommandlockouts
\usepackage{cite}
\usepackage{amsmath,amssymb,amsfonts}
\usepackage{algorithmic}
\usepackage{graphicx}
\usepackage{textcomp}
\usepackage{xcolor}
\def\BibTeX{{\rm B\kern-.05em{\sc i\kern-.025em b}\kern-.08em
    T\kern-.1667em\lower.7ex\hbox{E}\kern-.125emX}}
\ifCLASSOPTIONcompsoc
\usepackage[caption=false,font=normalsize,labelfon
t=sf,textfont=sf]{subfig}
\else
\usepackage[caption=false,font=footnotesize]{subfi
g}
\usepackage{xspace}
\fi
\begin{document}

\newcommand{\name}{mmHRR\xspace}
\newcommand\icdcs[1]{\color{black}{#1}\color{black}}

\title{\name: Monitoring Heart Rate Recovery with Millimeter Wave Radar
}

\author{
    \IEEEauthorblockN{
        Ziheng Mao\textsuperscript{1},
        Yuan He\textsuperscript{1\dag},\thanks{\textsuperscript{\dag} Yuan He is the corresponding author.}
        Jia Zhang\textsuperscript{1}, 
        Yimiao Sun\textsuperscript{1}, 
        Yadong Xie\textsuperscript{1}, 
        Xiuzhen Guo\textsuperscript{2}
    }
    \IEEEauthorblockA{
        \textsuperscript{1}School of Software \& BNRist, Tsinghua University, China\\
        \textsuperscript{2}College of Control Science and Engineering, Zhejiang University, China\\
        mzh23@mails.tsinghua.edu.cn, 
        heyuan@tsinghua.edu.cn, 
        \{j-zhang19, sym21\}@mails.tsinghua.edu.cn, \\
        ydxie@tsinghua.edu.cn, 
        guoxz@zju.edu.cn
    }
}

\maketitle

\begin{abstract}
Heart rate recovery (HRR) within the initial minute following exercise is a widely utilized metric for assessing cardiac autonomic function in individuals and predicting mortality risk in patients with cardiovascular disease. However, prevailing solutions for HRR monitoring typically involve the use of specialized medical equipment or contact wearable sensors, resulting in high costs and poor user experience. In this paper, we propose a contactless HRR monitoring technique, \name, which achieves accurate heart rate (HR) estimation with a commercial mmWave radar. Unlike HR estimation at rest, the HR varies quickly after exercise and the heartbeat signal entangles with the respiration harmonics. To overcome these hurdles and effectively estimate the HR from the weak and non-stationary heartbeat signal, we propose a novel signal processing pipeline, including dynamic target tracking, adaptive heartbeat signal extraction, and accurate HR estimation with composite sliding windows. Real-world experiments demonstrate that \name exhibits exceptional robustness across diverse environmental conditions, and achieves an average HR estimation error of 3.31 bpm (beats per minute), 71\% lower than that of the state-of-the-art method.
\end{abstract}


\input{chapter/1-introduction}
\input{chapter/3-design}
\input{chapter/4-evaluation}
\input{chapter/5-relatedwork}
\input{chapter/6-conclusion}

\section*{Acknowledgement}
This work is supported in part by the National Natural Science Foundation of China (No.
62425207 and No. U21B2007) and the Postdoctoral Fellowship Program of CPSF under Grant Number GZB20240356.

\bibliographystyle{IEEEtran}
\bibliography{ref.bib}

\end{document}

%% file: chapter/1-introduction.tex
\section{Introduction\label{intro}}

Nowadays, cardiovascular disease (CVD) is the leading cause of human mortality worldwide. In China, two out of every five deaths are attributed to CVD, while the number of CVD patients stands at approximately 350 million\cite{sheng2023report}. Cardiac autonomic dysfunction seems to be related to the occurrence of CVD and can lead to an increased risk of mortality in CVD patients. Earlier studies\cite{peccanha2014heart} have demonstrated that heart rate recovery (HRR), i.e., the decrease in heart rate (HR) after exercise, can serve as a practical and convenient method for assessing cardiac autonomic function in both CVD patients and healthy individuals. Furthermore, HRR has emerged as a prognostic indicator of mortality among high-risk patients\cite{cole1999heart}, which has been widely used in clinical practice.

\begin{figure}[!t]
\centering
\includegraphics[width=0.8\linewidth]
{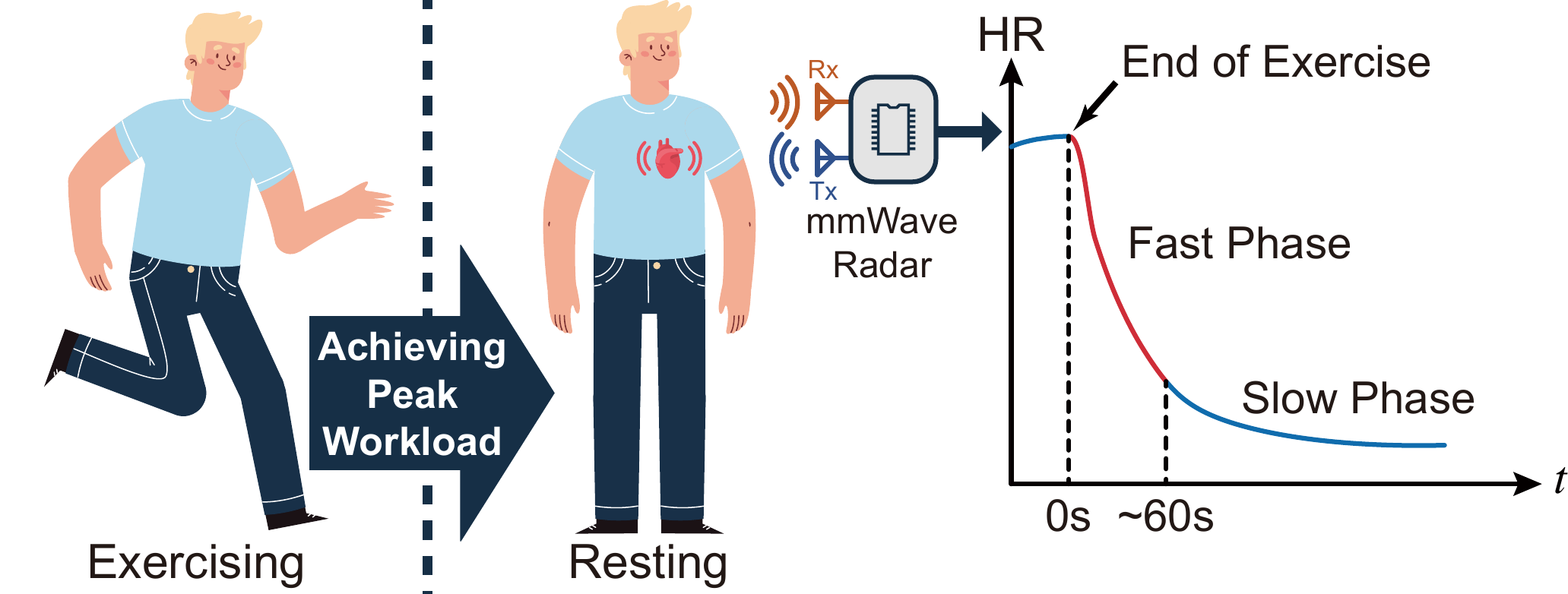}
\vspace{-0.2cm}
\caption{\name utilizes a mmWave radar to monitor heart rate recovery during the initial minute following the cessation of exercise.}
\label{fig:scenario}
\vspace{-0.5cm}
\end{figure}

Existing works show that the recovery of HR after exercise can be approximated as an exponential curve\cite{peccanha2014heart}, consisting of two parts, namely the fast recovery phase and the slow recovery phase. The former encompasses the first minute after exercise, during which HR decreases rapidly. The latter encompasses the period extending beyond the first minute until recovering to the resting HR. In practice, HRR can be defined as the curve or absolute value of the decrease in HR within 10 sec, 30 sec, 1 min, or longer after exercise. In this paper, we consider the HR decrease within 1 min as HRR for the demand for accuracy and convenience in daily life. Our methodology necessitates the user to stay in place and rest for 1 min after achieving peak workload, during which the HR is continuously estimated and the results could be exported for further analysis, as shown in Fig. \ref{fig:scenario}.

The prevailing measurement methods for HRR typically rely on specialized medical equipment or wearable devices\cite{fletcher2001exercise}. However, the issues of cost and comfort limit their widespread application. Recent advancements in wireless sensing technology have paved the way for HR monitoring using wireless signals, such as WiFi\cite{liu2018monitoring,wang2017phasebeat}, UWB\cite{zheng2020v2ifi}, and acoustics\cite{wang2023df}. Most of the works require the user to remain still and relaxed, which means that the heartbeat signal is seldom affected by the respiratory signal and can be easily extracted and separated. This also indicates that the user maintains a relatively low and stable HR. However, these characteristics are difficult to achieve in our scenario.

Recently, there has been an increasing interest in estimating human heartbeat waveforms or HR with mmWave technology\cite{yang2016monitoring,ha2020contactless,chen2021movi}. The short wavelength and large bandwidth characteristics of the mmWave signal provide much finer motion sensing resolution, facilitating the detection of tiny displacement in the human chest caused by the heartbeat.

\begin{figure}[!t]
\centering 
\subfloat[The change of HR in 60 sec.]{\includegraphics[width=0.4\linewidth]{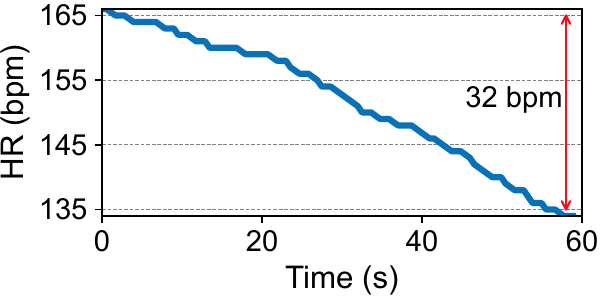}\label{fig:hrdrop}}
\hfil
\subfloat[Respiratory waveform.]{\includegraphics[width=0.4\linewidth]{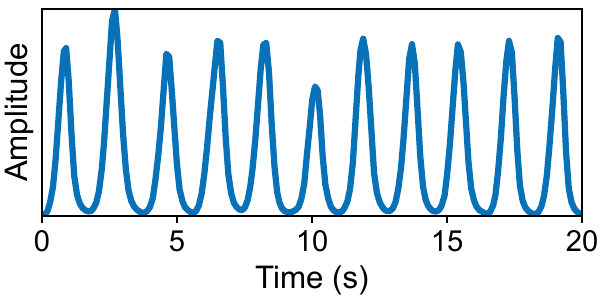}\label{fig:respiration}}
\vspace{0cm}
\subfloat[Frequency spectrum of the phase sequence.]{\includegraphics[width=0.8\linewidth]{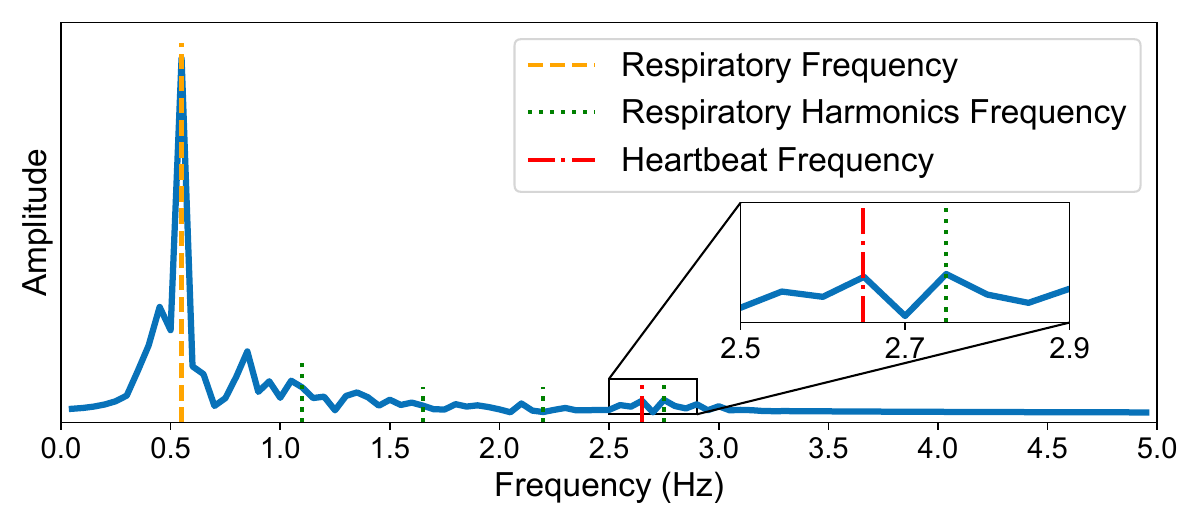}\label{fig:harmonic}}
\caption{Non-stationary property of heartbeat signal and interference of respiratory harmonics on heartbeat estimation.}
\vspace{-0.5cm}
\end{figure}

However, the direct migration of existing continuous HR monitoring methods for measuring HRR tends to result in significant estimation errors. On the one hand, the rapid decrease in HR complicates its continuous tracking.
We collect the HR of a volunteer within 60 sec after intense exercise through a commercial HR monitor\cite{PolarH10}. The result in Figure \ref{fig:hrdrop} shows a decrease in HR of 32 bpm.
Existing studies tend to employ Fourier analysis for HR estimation\cite{yang2016monitoring,wang2021mmhrv}. When HR decreases rapidly, the sampling window for the heartbeat signal needs to be narrowed to enable real-time tracking of HR. But a smaller window size substantially reduces the frequency resolution and leads to less accurate estimations.

On the other hand, the respiratory harmonics interfere with the heartbeat signal. As shown in Fig. \ref{fig:respiration}, the respiratory waveform deviates significantly from a sine wave, indicating the presence of several high-order harmonics in its Fourier expansion. We extract the chest motion signal from the radar reflection signal, its frequency spectrum is displayed in Figure \ref{fig:harmonic}. The frequencies of the harmonics often align closely with those of the heartbeat signal. Complicating matters further, during continuous changes in HR and respiratory rate (RR), the frequency of the heartbeat signal sometimes coincides with the frequencies of various respiratory harmonics multiple times, making it challenging to extract the heartbeat signal from the composite chest movements.

In this paper, we propose \name, which utilizes a commercial off-the-shelf (COTS) mmWave radar to estimate HRR from weak and non-stationary heartbeat signal disturbed by respiratory harmonics. To cope with the interference of respiratory harmonics, we model the chest movement, comprising the respiratory signal, its harmonics, and the heartbeat signal. We further design an adaptive process for extracting the heartbeat signal. The variational mode decomposition (VMD) algorithm is improved by incorporating the correlation coefficient and energy loss coefficient for parameter selection. Subsequently, the heartbeat signal is extracted based on the relationships between the frequencies of the respiratory harmonics. To continuously track the changes in HR, we devise a peak counting algorithm with composite sliding windows. By meticulously adjusting the size of these windows, this algorithm can accurately count the number of heartbeats in a certain window, thereby enabling real-time and precise estimation of HRR.

In summary, our main contributions are as follows:

\begin{itemize}
    \item We propose \name, a novel contactless HRR monitoring technique dedicated to estimating the human HR after exercise for assessing cardiac autonomic function.
    \item We design a tailored signal processing pipeline to achieve adaptive heartbeat signal extraction and accurate HR estimation from the weak and non-stationary heartbeat signals disturbed by respiratory harmonics.
    \item We conduct experiments with various settings in real scenarios. The results show that \name achieves an average estimation error of 3.31 bpm (beats per minute) in HR, 71\% lower than that of the state-of-the-art (SOTA) method.
\end{itemize}

The rest of the paper is organized as follows. Section \ref{design} elaborates on our design. Section \ref{evaluation} presents the implementation and evaluation results. Section \ref{relatedwork} discusses the related work. We conclude this work in Section \ref{conclusion}.

%% file: chapter/3-design.tex
\section{Design\label{design}}

\subsection{Overview}

\begin{figure*}[!t]
\centering
\includegraphics[width=0.85\linewidth]{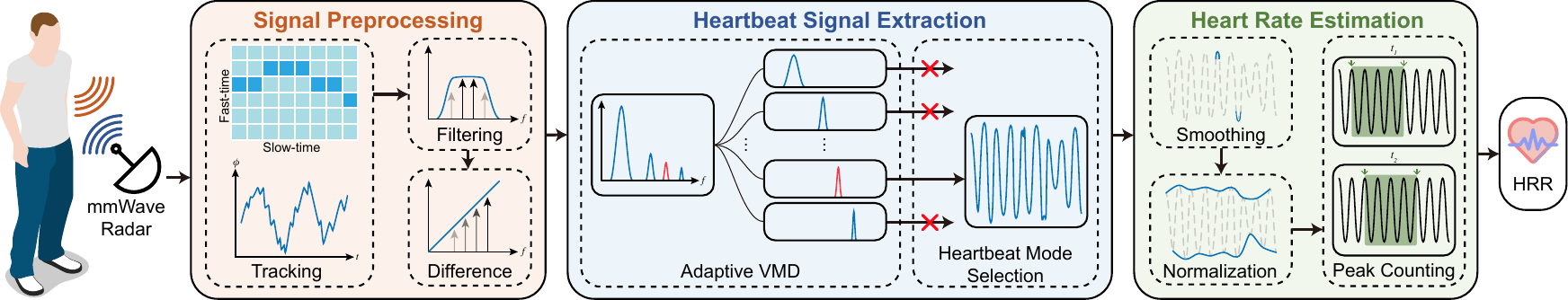}
\vspace{-0.2cm}
\caption{The architecture of \name.}
\label{fig:pipeline}
\vspace{-0.5cm}
\end{figure*}

We design a contactless HRR monitoring technique, \name, which aims to track the decrease of HR after exercise using COTS mmWave radar. As shown in Fig. \ref{fig:pipeline}, \name consists of the following three modules.

\textit{Signal Preprocessing}: We locate the target from the radar reflection signal and extract the chest motion signal. We apply a series of cascade filters to the signal to attenuate ambient noise and accentuate the frequency domain characteristics of the heartbeat signal.

\textit{Heartbeat Signal Extraction}: We model the chest motion signal including the compound movement of respiration and heartbeat, thus applying an adaptive VMD algorithm to separate the respiratory signal and its harmonics from the heartbeat signal. The algorithm allows for the effective handling of diverse scenes and fluctuating HR. After that, the heartbeat signal can be obtained by a mode selection algorithm.

\textit{Heart Rate Estimation}: We apply smoothing and amplitude normalization to the heartbeat signal and devise a peak counting algorithm based on adaptive-sized sliding window to accurately track the changes in the HR.

\subsection{Signal Preprocessing}

The mmWave frequency modulated continuous wave (FMCW) radar transmits chirps, whose frequencies increase linearly over time. The transmitted signal and the reflected signal are mixed obtain the intermediate frequency (IF) signal. A Fast Fourier transform (FFT) operation, known as Range-FFT\cite{iovescu2017fundamentals}, is applied to the IF signal. Each bin in the spectrum of IF signal, namely a range bin, corresponds to the intensity of the reflected signals from reflection points at various distances. Moreover, the displacement of a reflection point can be derived as $\Delta d=\frac{\lambda\Delta\phi}{4\pi}$, where $\lambda$ is the wavelength of the mmWave signal, and $\Delta\phi$ is the phase change of a range bin over time.\cite{guo2021measuring} Consequently, by identifying the peak of the range spectrum, we can obtain a phase sequence characterizing chest motion. The phase sequence, called the chest motion signal,  encompasses both respiratory and heartbeat movements.

Considering the inevitable body movement after exercise (e.g., the involuntary shaking of torso), the peak of the range spectrum may shift to the adjacent range bin over time, which results in a discontinuity in the phase sequence since the phase values in different range bins differ at the same moment. Fortunately, we find that only the relative displacement characterized by the phase difference needs to be extracted, allowing us to add or subtract a value from a phase sequence segment when the user switches to another range bin, thereby rendering the entire phase sequence continuous.

The extracted phase sequence is interfered with by ambient noise, making the separation of heartbeat signal more difficult. For an ordinary adult, the RR is generally higher than 10 Bpm\cite{yuan2013respiratory}, while the maximum HR is generally not more than 200 bpm\cite{tanaka2001age}. Therefore, we use a band-pass filter of 0.2-3.4 Hz to attenuate low and high-frequency noise.

In addition, it is worth noting that the amplitude of respiratory signals is typically several times larger than that of heartbeat signals. A difference operation is applied to the signal to facilitate the identification of each component during subsequent signal decomposition. This approach diminishes the energy of low-frequency respiratory components, while comparatively amplifying the energy of high-frequency heartbeat and respiratory harmonics components.

\subsection{Heartbeat Signal Extraction\label{extraction}}

In the previous section, we elaborate on how to use the phase sequence to characterize the movement of the human chest, which includes the displacement caused by human respiration and heartbeat. We construct a model to describe how these motion signals are superposed. Subsequently, an adaptive VMD algorithm is applied to decompose the chest motion signal into different modes\footnote{Note that the term \textit{modes} denotes the decomposed signals obtained from VMD, while the term \textit{components} refers to the inherent elements in the signal.}, and we design an algorithm to select the heartbeat signal among them.

\subsubsection{Chest Motion Mixture Model}

We model the chest motion signal, donated as $x(t)$, as a superposition of the respiratory-induced motion $x_r(t)$, the heartbeat-induced motion $x_h(t)$, and random noise $n(t)$. 
Both $x_r(t)$ and $x_h(t)$ are quasi-periodic real even signals, which can be expressed as Fourier series, considering $x_r(t)$ as an example:

\begin{equation}
\begin{aligned}
& x_r(t) = \frac{a_{r0}}{2} + \sum_{n=1}^{\infty} a_{rn} \cos(n\omega t), \\
& a_{r0} = \frac{2}{T} \int_{0}^{T} x(t)dt,\quad a_{rn} = \frac{2}{T} \int_{0}^{T} x(t) \cos(n\omega t)dt,
\end{aligned}
\end{equation}

\noindent where the direct current (DC) component $a_{r0}$ has been removed by band-pass filtering. Since the respiratory signal is not exactly sinusoidal, it results in the presence of high-order harmonics in its Fourier series representation. During the recovery of HR, the frequency of the heartbeat signal will be close to or even coincide with the frequencies of harmonics, thereby posing a significant challenge in their separation.

\begin{figure}[!t]
\centering
\includegraphics[width=0.9\linewidth]{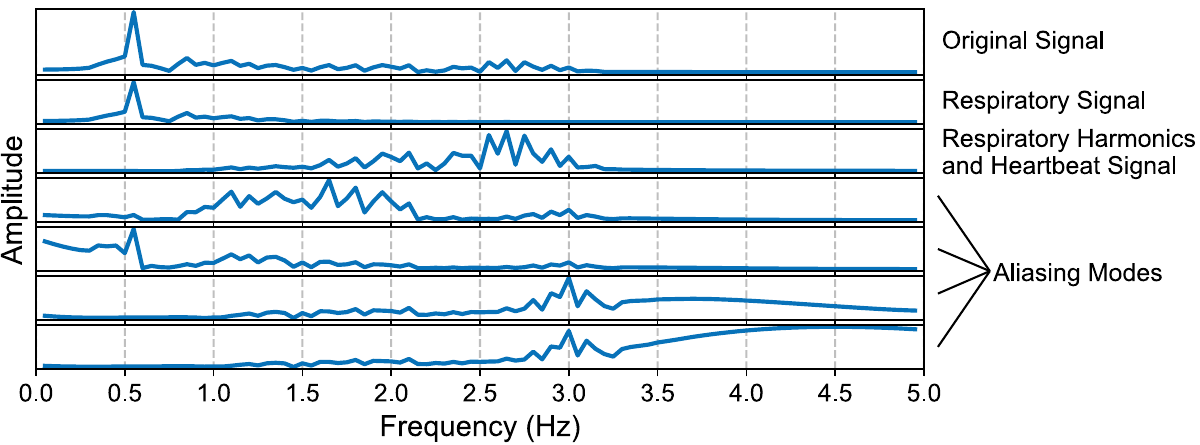}
\vspace{-0.2cm}
\caption{The decomposition result when encountering mode aliasing.}
\label{fig:dec_result1}
\vspace{-0.5cm}
\end{figure}

\begin{figure*}[!t]
\centering
\includegraphics[width=0.9\linewidth]{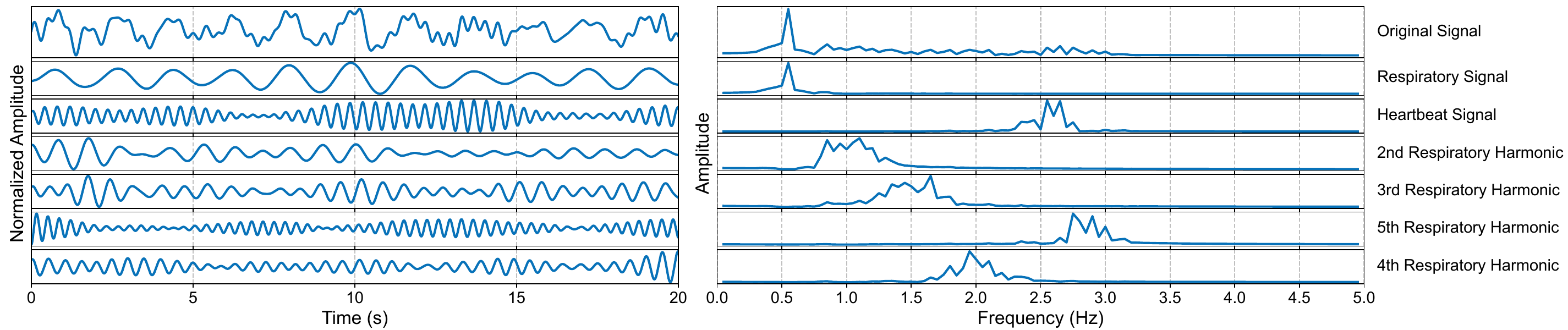}
\vspace{-0.2cm}
\caption{The heartbeat signal and respiratory harmonics in the VMD decomposition result.}
\label{fig:dec_result}
\vspace{-0.5cm}
\end{figure*}

\subsubsection{Separating Heartbeat Signal}

To separate the heartbeat signal from the interference mentioned above, we utilize the VMD algorithm\cite{dragomiretskiy2013variational} to decompose the chest motion signal into different modes. This algorithm employs a variational approach to decompose the original signal into $K$ narrow-band signals $u_k$ with center frequencies $\omega_k$. The VMD algorithm is capable of extracting multiple dominant components from the chest motion signal, thus facilitating the extraction of the heartbeat signal.

In the practical application of the VMD algorithm, the main challenge lies in the parameter selection. The two key parameters, the number of components $K$ and the penalty factor $\alpha$, can greatly affect the performance of the decomposition. Specifically, when $K$ is too large, the decomposition results may include noise modes, whereas setting it too small may result in the loss of meaningful components. Given that the energies of the sixth or higher harmonics of the respiratory signal, as well as the harmonics of the heartbeat signal, are so weak as to be practically negligible, it can be inferred that there are at most 6 meaningful components in $x(t)$. Typically, random noise possessing very low energy would fall into the residual of the VMD result. However, noise modes may emerge if there are insufficient meaningful components. In such instances, these noise modes are removed in the subsequent \textit{Heartbeat Mode Selection} section, ensuring they do not compromise the accuracy of HRR estimation.

The penalty factor $\alpha$ influences the bandwidth of each mode. A smaller value of $\alpha$ results in a wider bandwidth, which can lead to mode aliasing, where two components or a component and noise are combined into a single mode. Conversely, a larger value of $\alpha$ imposes a tighter constraint on the bandwidth of each mode, potentially causing over-decomposition, where a single component may be decomposed into multiple modes.

To solve the problems, we use two coefficients to determine the permissible value range of $\alpha$. First, we evaluate the occurrence of mode aliasing by measuring the similarity between modes. As shown in Fig. \ref{fig:dec_result1}, when the bandwidth constraint is too loose, the modes will cover a wider frequency range, resulting in more overlap between different modes, thereby increasing the similarity. We calculate the Pearson correlation coefficient between each pair of modes as $r_{ij}=\frac{E(u_iu_j)-E(u_i)E(u_j)}{\sqrt{D(u_i)D(u_j)}}$, where $E(\cdot)$ and $D(\cdot)$ denote the mean and variance, respectively. The value of $r_{max}$ exceeding $\mu_1$ (the empirical value of $\mu_1$ is set to 0.2) implies the presence of highly similar modes. Consequently, the coefficient limits the lower bound of $\alpha$.

Then, the energy loss coefficient is introduced to quantify the extent of information loss incurred throughout the decomposition. The energy loss coefficient is defined as the ratio of the decomposition residual energy to the original signal energy, and can be calculated as $p=\frac{\left\|f-\sum u_{k}\right\|_{2}^{2}}{\|f\|_{2}^{2}}$, where $f$ denotes the original signal. As $\alpha$ increases, there is a corresponding incremental elevation in the energy loss attributable to decomposition. By limiting $p$ to not exceed $\mu_2$ (empirically set to $10^{-4}$), the upper bound of $\alpha$ can be restricted to avoid over-decomposition.

These two stipulations collectively define a permissible range of values for $\alpha$. However, it is not feasible to directly derive the range from the provided values of $r_{max}$ and $p$. One potential method to estimate the range involves enumerating $\alpha$, which is time-consuming. Since any value within the range is capable of yielding effective decomposition results, an arbitrary value of $\alpha$ within the range can be efficiently determined through binary search. This value is then employed as the input for the VMD algorithm.

As discussed in Section \ref{intro}, the frequency of the heartbeat signal changes rapidly after exercise. Nonetheless, the VMD algorithm assumes that each mode is a narrow-band signal. When a relatively long segment of the signal is decomposed, the resulting modes exhibit slight frequency variation and fail to capture the non-stationary property of the heartbeat signal, leading to significant errors in HR estimation. We propose a sliding window approach, where the signal is divided into shorter segments, and the VMD algorithm is applied within each window. The specific design of the sliding window is coupled with the design of \textit{Heart Rate Estimation} in the next section and will be described in detail later.

\subsubsection{Heartbeat Mode Selection}

Fig. \ref{fig:dec_result} presents a visual representation of the VMD decomposition result, with the modes sorted in descending order of energy. The decomposition results reveal the presence of the respiratory signal, its four harmonics, and the heartbeat signal distributed among the $K$ modes obtained from the VMD algorithm. However, due to the inability to anticipate the relative energy of the heartbeat signal and the respiratory harmonics, direct selection of the heartbeat signal from the $K$ modes becomes challenging. Therefore, a dedicated method must be devised to accurately identify the mode corresponding to the heartbeat signal.

The identification of the heartbeat signal is achieved through the analysis of the frequency domain characteristics of the modes. Initially, the noise signals, characterized by a broad bandwidth and insignificant frequency peaks, are filtered out. Subsequently, the remaining modes are sorted by their peak frequencies. The energy of the respiratory signal demonstrates the highest magnitude, hence it can be easily identified. The next step involves distinguishing the heartbeat signal from the other modes. Typically, modes whose peak frequencies are integer multiples of that of the respiratory signal are regarded as respiratory harmonics and are consequently excluded. If a discernible frequency gap exists between the frequencies of the heartbeat signal and the respiratory harmonics, we can simply select the one exhibiting the highest frequency peak among the remaining modes as the heartbeat signal.

In certain instances, the frequency of the heartbeat signal may coincide precisely with that of an integer multiple of the respiratory frequency. Consequently, in the frequency spectrum, the heartbeat signal overlaps with the respiratory harmonic. In this case, we choose the respiratory harmonic with the highest energy as the heartbeat signal.

\begin{figure}[!t]
\centering
\includegraphics[width=0.8\linewidth]{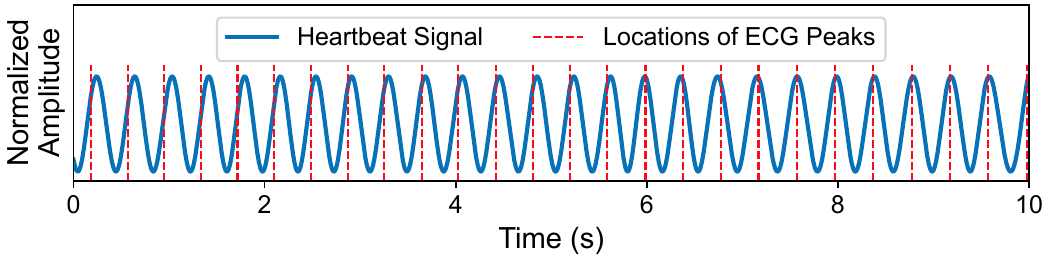}
\vspace{-0.3cm}
\caption{Comparison between the processed heartbeat signal and the ground truth.}
\label{fig:time_comp}
\vspace{-0.5cm}
\end{figure}

\subsection{Heart Rate Estimation\label{estimation}}

Based on the heartbeat signals selected in the previous section, we proceed to estimate the HR of the target.

Considering the limitations associated with Fourier analysis, we design a peak counting algorithm to estimate HR from the time domain characteristics of the heartbeat signal. In order to mitigate the variable amplitude and small oscillations present in the signal, which can significantly impact the accuracy of peak counting, we employ a preprocessing approach. First, a moving average filter is applied to reduce the high-frequency noise and eliminate small fluctuations. Then, we divide the original signal by its envelope to normalize the amplitude in the time domain. Through these preprocessing steps, the periodic characteristics of the signal are significantly enhanced. Fig. \ref{fig:time_comp} shows a comparison between the processed heartbeat signal and the ground truth, where the dashed lines indicate the position of each peak in the ECG. Following amplitude normalization, a \textit{peak} of the heartbeat signal is characterized as a local maximum exhibiting an amplitude of no less than 0.5, with a minimum interval of 0.27 sec between two consecutive occurrences (equivalent to a maximum HR of 220 bpm). Notably, the peaks of the processed heartbeat signal align closely with those of the ECG, affirming the effectiveness of the proposed approach.

\begin{figure}[!t]
\centering
\includegraphics[width=0.9\linewidth]{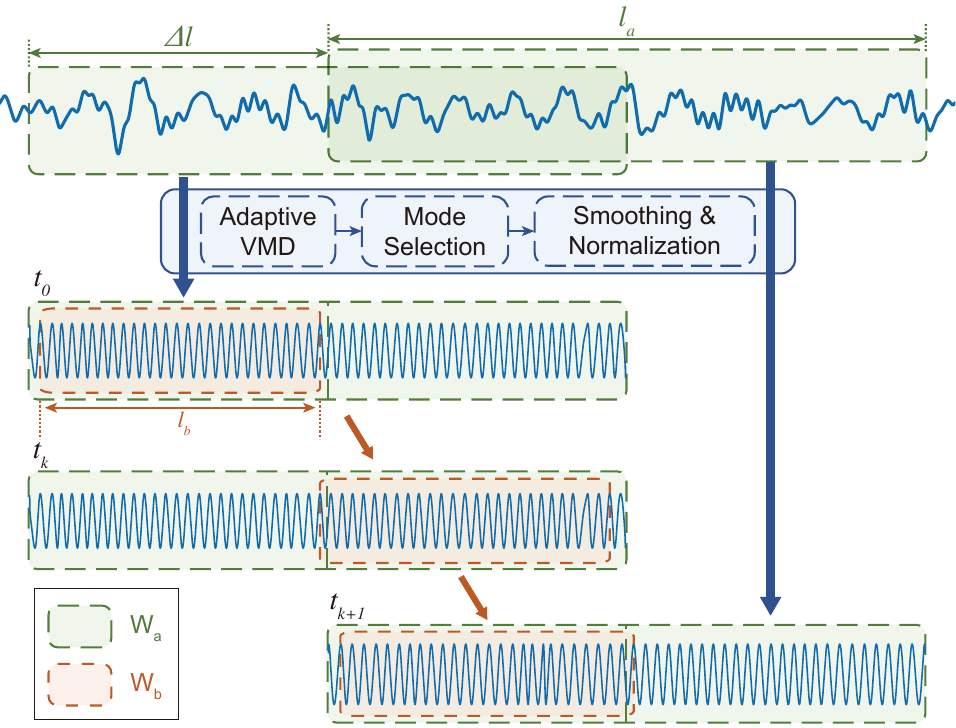}
\vspace{-0.2cm}
\caption{The composite sliding windows algorithm of \name.}
\label{fig:sliding}
\vspace{-0.5cm}
\end{figure}

To account for both time resolution and statistical result precision, we further design a peak counting method based on variable-sized sliding windows. These windows are positioned based on the peak values in the heartbeat signal, with the left and right endpoints of each window coinciding with these peaks. As we move the window forward over time, we ensure that the window size $l_b$ does not fall below a predetermined threshold $l_{min}$. We finally estimate HR by dividing the number of peaks within the window by the window length.

To accommodate HR variations, we propose a dynamic adjustment of the threshold value $l_{min}$. When the HR is high, $l_{min}$ can be as to a smaller value. Conversely, a larger $l_{min}$ can be used. To implement this adaptive approach, we initially set $l_{min}$ to a predetermined value and estimate the approximate HR using the previously discussed peak counting algorithm. Subsequently, we re-execute the peak counting algorithm and dynamically adjust $l_{min}$ based on the estimated HR within a defined range during the sliding window process. This iterative adjustment enhances the accuracy of HR estimation by accounting for the changes in HR.

In Section \ref{extraction}, we outline the necessity of segmenting the chest motion signal into smaller segments before applying the VMD algorithm. Here, we denote this segmentation window as $W_a$. To incorporate $W_a$ into the sliding window of the peak counting algorithm, denoted as $W_b$, we create composite sliding windows that encompass both $W_a$ and $W_b$.  Assuming a fixed size of $W_a$ as $l_a$, we slide $W_a$ forward by a distance of $\Delta l$ each time. After every $W_a$ slide, the VMD is applied to the chest motion signal contained within the window. We set $\Delta l$ as the maximum possible value of $l_b$, while $l_a$ is set as twice the maximum possible value of $l_b$. During the sliding of $W_b$, if its left endpoint falls within the range of next $W_a$, $W_a$ slides forward. This approach guarantees that $W_a$ always contains $W_b$ and minimizes its length.
Fig. \ref{fig:sliding} illustrates the composite sliding windows. At $t_0$, $W_b$ is located at the beginning of the first $W_a$. After several slides of $W_b$, $W_b$ approaches the midline of $W_a$ at $t_k$. At the next moment $t_{k+1}$, the left endpoint of $W_b$ falls within the range of the next $W_a$, and thus $W_a$ slides forward by a distance $\Delta l$. In each moment, we count the number of peaks in $W_b$ and divide it by $l_b$ to calculate the HR.

%% file: chapter/4-evaluation.tex
\section{Evaluation\label{evaluation}}

\begin{figure}[!t]
\centering 
\includegraphics[width=\linewidth]{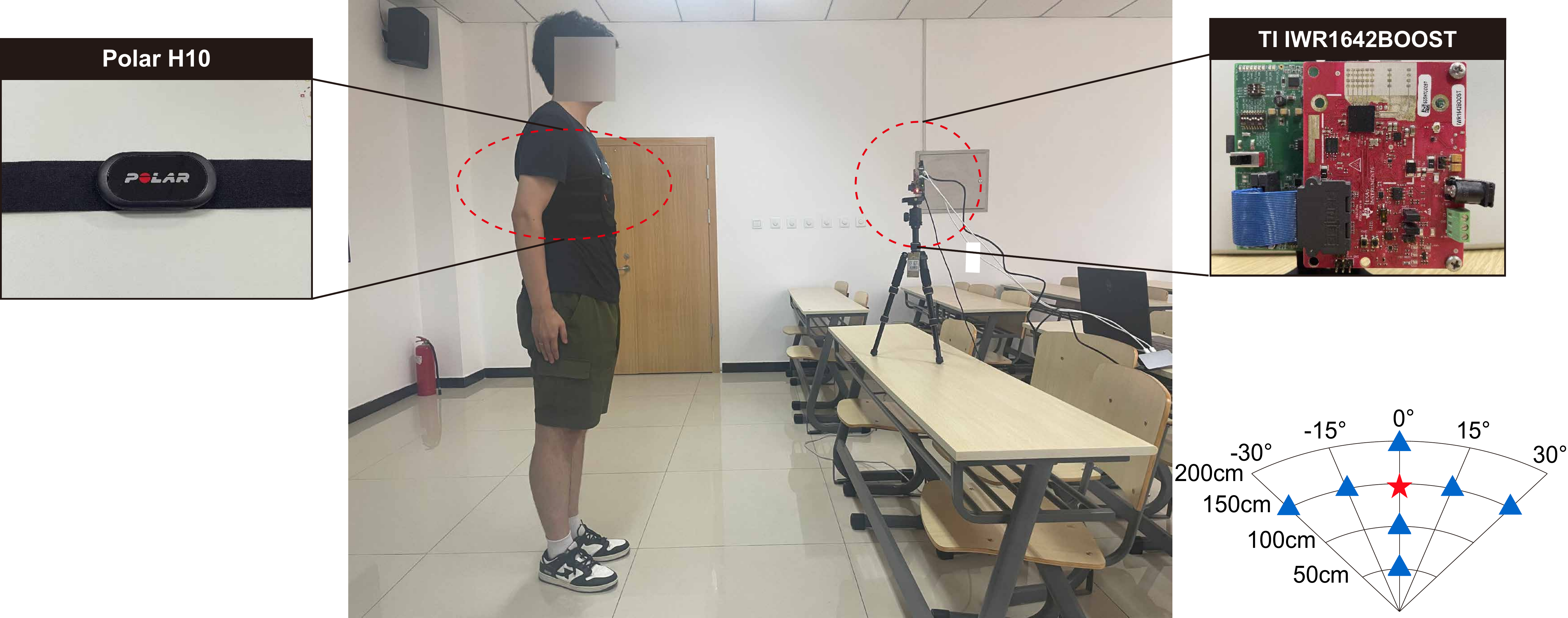}
\vspace{-0.6cm}
\caption{Implementation and deployment of \name.\label{fig:settings}}
\vspace{-0.5cm}
\end{figure}

\begin{figure*}[!t]
\centering
\begin{minipage}[t]{0.24\linewidth}
\centering
\includegraphics[width=\linewidth]{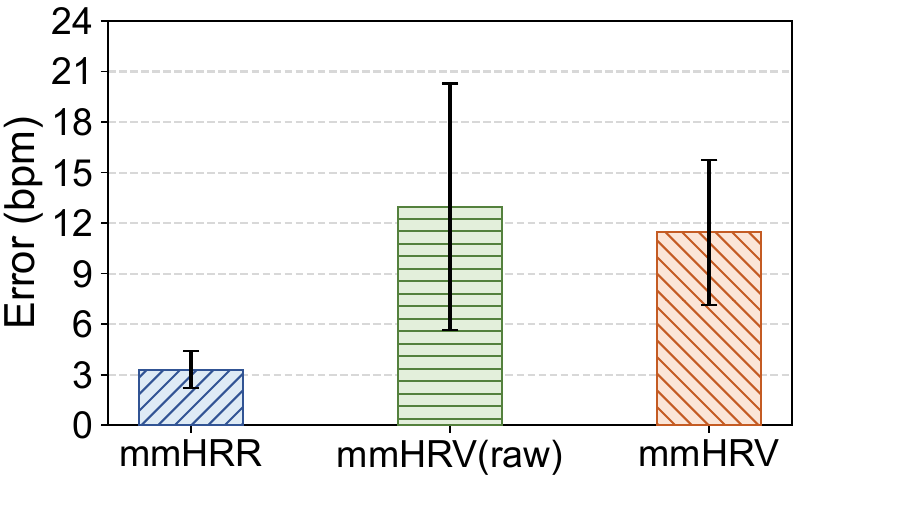}
\vspace{-0.6cm}
\caption{Overall performance.\label{fig:overall}}
\end{minipage}
\hspace{0.5cm}
\begin{minipage}[t]{0.24\linewidth}
\centering
\includegraphics[width=\linewidth]{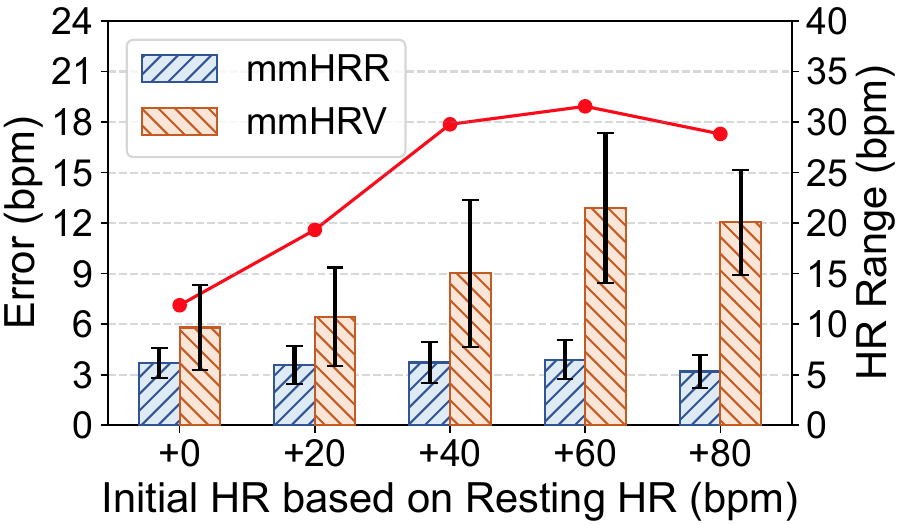}
\vspace{-0.6cm}
\caption{Performance at different initial HRs.\label{fig:inithr}}
\end{minipage}
\hspace{0.5cm}
\begin{minipage}[t]{0.24\linewidth}
\centering
\includegraphics[width=\linewidth]{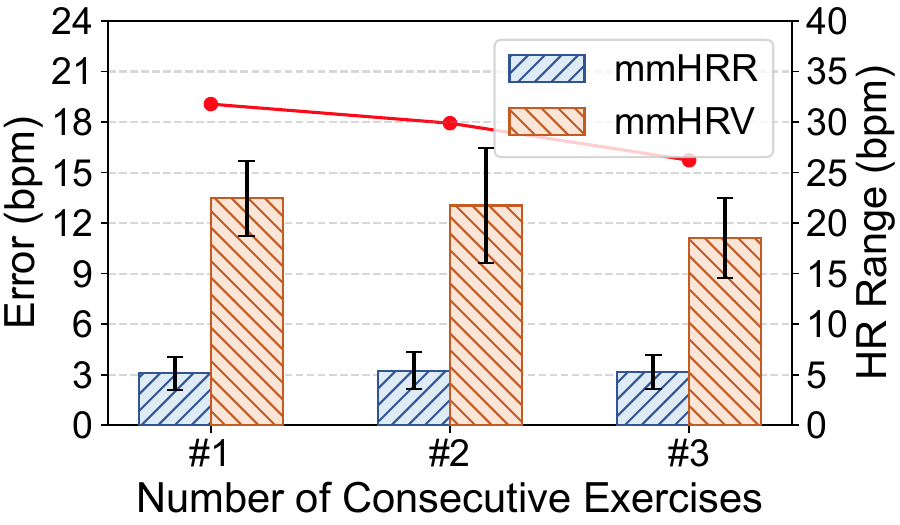}
\vspace{-0.6cm}
\caption{Performance under consecutive exercises.\label{fig:threetimes}}
\end{minipage}
\vspace{-0.3cm}
\end{figure*}

\begin{figure*}[!t]
\centering
\begin{minipage}[t]{0.24\linewidth}
\centering
\includegraphics[width=\linewidth]{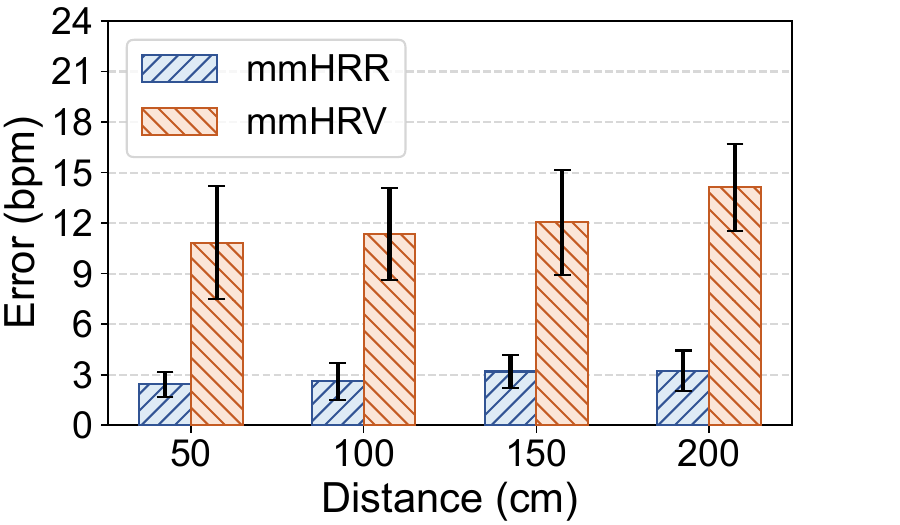}
\vspace{-0.6cm}
\caption{Performance at different distances.\label{fig:distance}}
\end{minipage}
\begin{minipage}[t]{0.24\linewidth}
\centering
\includegraphics[width=\linewidth]{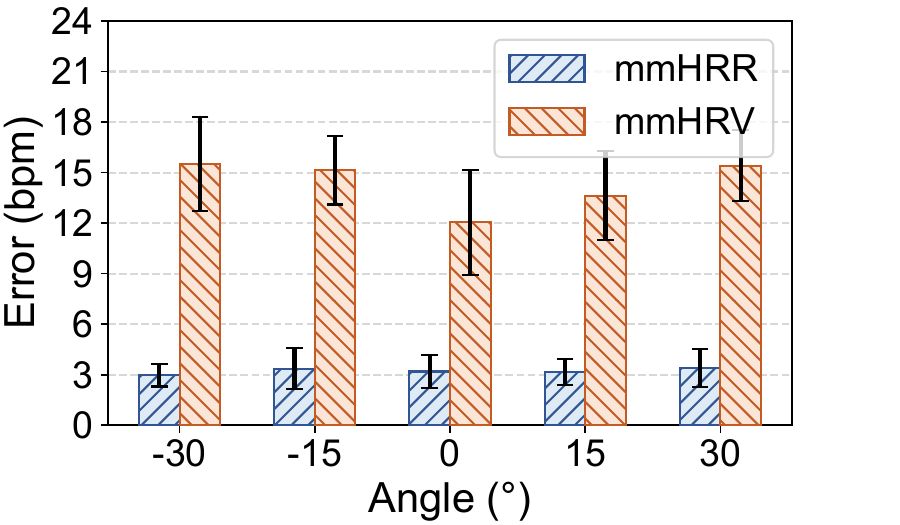}
\vspace{-0.6cm}
\caption{Performance at different angles.\label{fig:angle}}
\end{minipage}
\begin{minipage}[t]{0.48\linewidth}
\centering
\includegraphics[width=\linewidth]{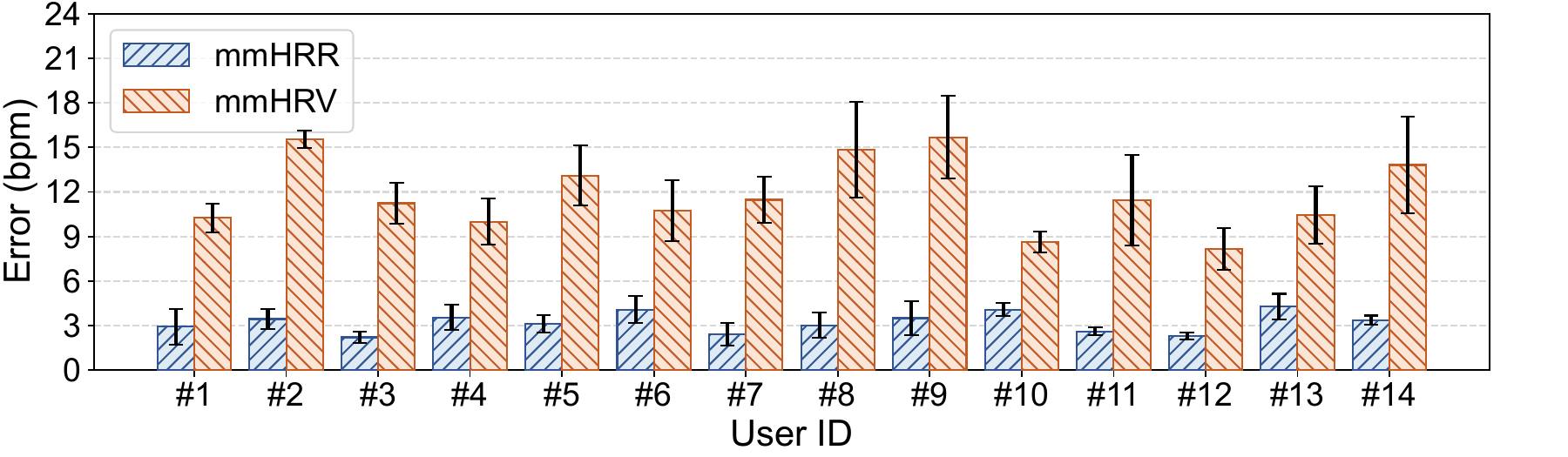}
\vspace{-0.6cm}
\caption{Performance for different users.\label{fig:users}}
\end{minipage}
\vspace{-0.5cm}
\end{figure*}

\subsection{Implementation}

We implement \name on a commercial mmWave radar, TI IWR1642BOOST \cite{IWR1642}. The radar operates in the frequency band of 77-81 GHz and is equipped with 2 transmitting antennas and 4 receiving antennas. The raw data is captured by a TI DCA1000EVM capture board\cite{DCA1000}, and forwarded over Gigabit Ethernet to PC. We implement our data processing pipeline in Python, which runs on a laptop with an Intel Core i7-9750H CPU and 16 GB memory.

\subsection{Methodology}

We invite 14 healthy users, comprising 11 males and 3 females, to conduct the experiment. Their heights range from 165 cm to 190 cm, and their weights range from 55 kg to 90 kg. The experiments are conducted following the IRB protocol of our institute. Before each experiment, a user are given sufficient rest and then perform exercise activities until reaching a predetermined HR (referred to as the initial HR). Subsequently, the user immediately stand in place, and the mmWave radar is activated to collect data within 60 sec. The experiments take place in a classroom, as shown in Fig. \ref{fig:settings}, where different target positions are marked in the corner. 
Experiments with different initial HRs are conducted when users are standing at the position marked by the red star, while the initial HR remains fixed in other positions. The ground truth of HR is captured by a commercial sensor, Polar H10 \cite{PolarH10}. It is connected to our PC via Bluetooth and is synchronized with the radar through the local timestamp.

We utilize the absolute HR estimation error as the metric to evaluate the performance of \name. The error is defined as the absolute difference between the ground truth $HR_G$ and the estimated $HR_E$, i.e., $\Delta HR = \left|HR_G-HR_E\right|$.

We implement a SOTA work mmHRV\cite{wang2021mmhrv} as the baseline method. mmHRV applies a VMD-like algorithm to separate the interference of respiration and other clutters from the heartbeat signal. Such a method has been claimed to achieve accurate HRV estimation when the human target is sitting still, Specifically, an average absolute interbeat interval (IBI) estimation error of 28 ms is achieved. Since we can calculate the HR from the IBI by $HR=60/IBI$, the estimated HR can be obtained through mmHRV.

\subsection{Overall Performance}

We first evaluate the overall performance of \name. As shown in Fig. \ref{fig:overall}, the average HR estimation error of \name is 3.31 bpm with a standard deviation of 1.08 bpm. For comparison, the error of mmHRV is 12.96 bpm with a standard deviation of 7.33 bpm (referred to as mmHRV(raw) in the figure). We notice that mmHRV employs a heuristic algorithm to select the parameters of the VMD algorithm and to choose the mode corresponding to the heartbeat signal. This algorithm overlooks the interference from the respiratory harmonics, therefore often misidentifies the heartbeat signal, resulting in significant estimation errors. When excluding such specific data, the average error of mmHRV decreases to 11.45 bpm. The performance of mmHRV is mainly limited by the lack of signal preprocessing and a suboptimal selection of VMD parameters. In the following comparisons, we adopt the modified mmHRV method as our baseline.

\begin{figure*}[!t]
\centering
\begin{minipage}[t]{0.24\linewidth}
\centering
\includegraphics[width=\linewidth]{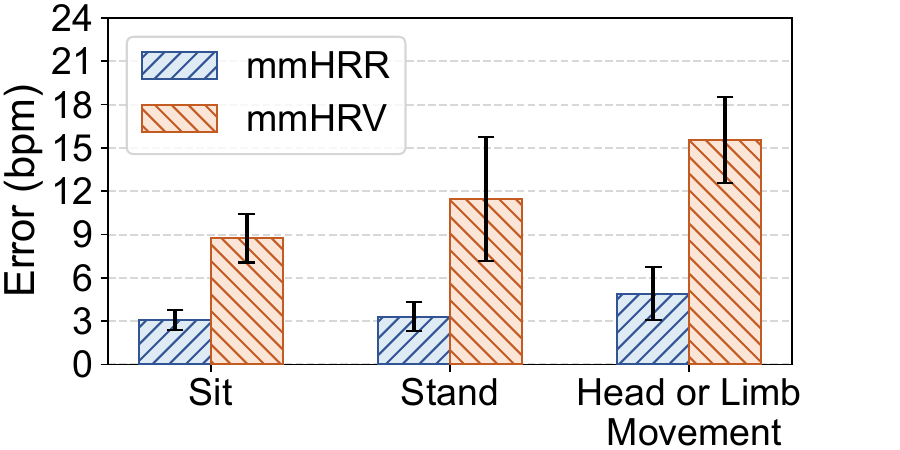}
\vspace{-0.7cm}
\caption{Performance under different user postures.\label{fig:posture}}
\end{minipage}
\hspace{0.5cm}
\begin{minipage}[t]{0.24\linewidth}
\centering
\includegraphics[width=\linewidth]{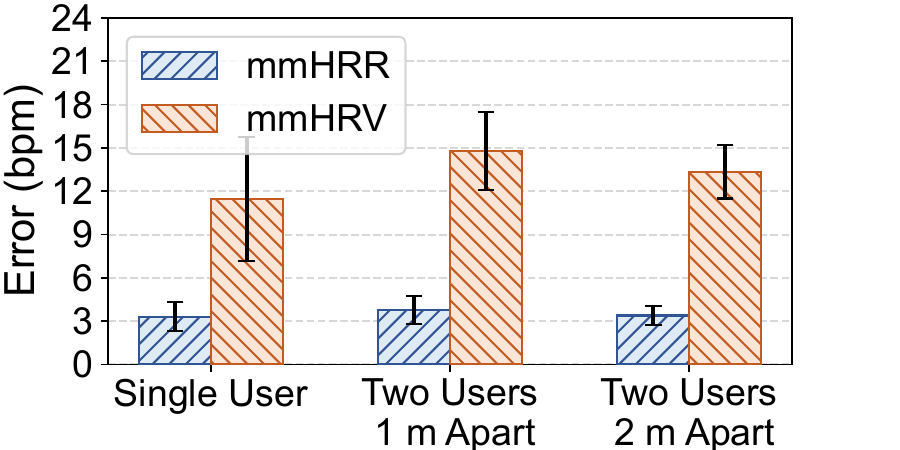}
\vspace{-0.7cm}
\caption{Performance with multiple users.\label{fig:multi}}
\end{minipage}
\hspace{0.5cm}
\begin{minipage}[t]{0.24\linewidth}
\centering
\includegraphics[width=\linewidth]{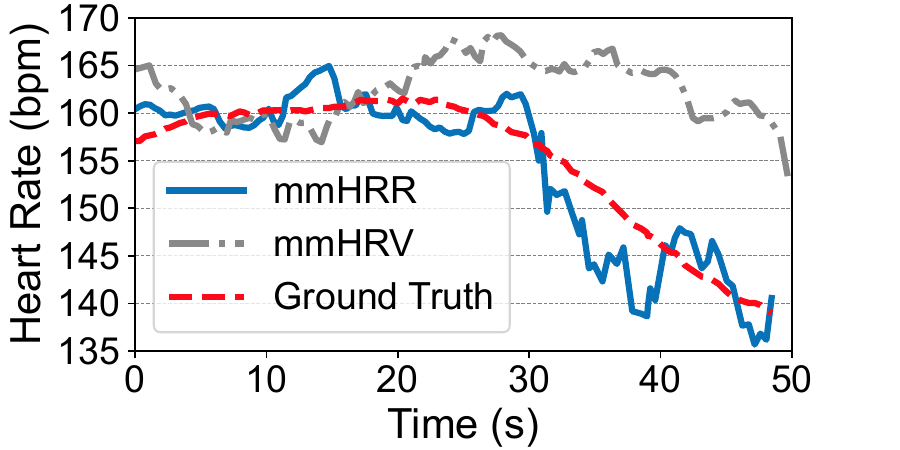}
\vspace{-0.7cm}
\caption{The estimation result when different types of HR variations occur.\label{fig:inc_hr}}
\end{minipage}
\vspace{0cm}
\vspace{-0.5cm}
\end{figure*}

\subsection{Impacting Factors}


\subsubsection{Initial HR} We begin by evaluating the impact of different initial HRs on the performance of \name. Fig. \ref{fig:inithr} shows the average estimation errors of \name and mmHRV for 5 different initial HRs, where +x indicates that the initial HR is x bpm higher than the resting HR of the user. The figure also shows the relationship between the decreasing rate of HR (calculated from the ground truth HR data) and the initial HR. The decreasing rate is determined as the HR range within a fixed period (i.e., 60 sec) and is represented by the red line in the figure. The HR range basically increases with the increase of initial HR but decreases at +80 bpm. The reason is that sometimes the HR  remains constant for a while before it starts to decrease rapidly. The results show that mmHRV achieves the smallest average estimation error of 5.8 bpm when users are at the resting HR. It is worth noting that the average estimation error of mmHRV does not strictly increase in proportion to the initial HR but is positively correlated with the HR range. The reason is that a higher HR range corresponds to a faster HR change, which can lead to the degradation of the VMD algorithm. However, the performance of \name remains relatively unaffected by this factor, which is primarily due to the adaptive sliding window method. Our approach has better performance at +80 bpm due to the weaker energy of the respiratory harmonics and a lesser effect on the heartbeat signal.

\subsubsection{HR Decreasing Rate} We delve into the impact of HR decreasing rate on performance. We ask each user to perform three consecutive experiments with the highest initial HR to observe the changes in the estimation error. Fig. \ref{fig:threetimes} shows the average HR estimation error of \name and mmHRV, as well as the HR range, across the three consecutive sets of experiments. During multiple intense exercises, the decreasing rate of HR gradually slows down, thereby facilitating the estimation of HR. This explains the slight decrease in the estimation error of mmHRV. In contrast, the performance of \name remains consistent and outperforms mmHRV throughout the experiments. These results underscore the efficacy of \name in monitoring HRR across multiple exercise sessions.

\subsubsection{Distance Between the User and the Radar} We fix the initial HR of the users to +80 bpm and evaluate the effect of the distance between the user and the radar on the performance of \name. Fig. \ref{fig:distance} shows the average estimation error of \name and mmHRV at 4 distance points within [50, 200] cm. The analysis reveals that as the propagation distance of the mmWave signals increases, the reflected signal becomes more noisy, leading to increasing average estimation errors. However, the performance of \name is relatively unaffected by the distance. Even at the farthest distance of 200 cm, the average HR estimation error remains below 3.24 bpm. This superiority can be attributed to the effectiveness of our signal preprocessing pipeline in handling noisy reflected signals and accurately extracting the heartbeat-related phase sequence.

\subsubsection{The Angle of the User Relative to the Radar} We evaluate the performance when the user is not directly in front of the radar. Fig. \ref{fig:angle} shows the average estimation error of \name and mmHRV over different angles within [-30°, 30°]. In general, due to the directivity of the antenna, the average estimation error rises as the user deviates from the direct front position. The best performance of mmHRV is achieved when the user is directly in front of the radar. In contrast, the variation in the estimation error of \name with respect to the angle remains minimal. This can be credited to the robustness of our tracking and denoising process in effectively mitigating the adverse impacts of low signal-to-noise ratio (SNR) signals. In summary, The results demonstrate the capability of \name to reliably estimate HRR in indoor scenarios.

\subsubsection{User Heterogeneity} Fig. \ref{fig:users} shows the average estimation error for \name and mmHRV across 14 different users. \name exhibits different estimation errors for different users, with average errors ranging from 2.21 to 4.28 bpm. This discrepancy can be attributed to the distinct physical characteristics of the users, including their height, weight, RR, HR, and amplitude of respiration. Furthermore, \name consistently outperforms mmHRV across all users, which can be largely attributed to our specialized heartbeat signal extraction and HR estimation algorithms.

\subsection{Robustness study}


\subsubsection{User Postures} We assess \name's robustness in response to different user postures by conducting additional experiments. In these experiments, users are instructed to either sit or to move their head and limbs slightly while standing, simulating the user's relaxation after exercise. The results of these experiments are shown in Fig. \ref{fig:posture}. The average estimation error decreases to 3.09 bpm when users are seated, which can be attributed to a significant reduction in torso movement. Conversely, when users engage in activity, there is an increase in the average estimation error to 4.9 bpm. This increase can be attributed to the fact that the movement of limbs or head causes movement of their chest and a lower SNR of the received signal.

\subsubsection{Multiple Users} We measure the ability of \name to simultaneously measure the HRR of multiple users. Two users are asked to stand at different distances from the radar, each at a distinct angle relative to the radar to prevent mutual block. The distance between the users is set as 1 m and 2 m, respectively. We enhance the tracking algorithm, enabling it to monitor targets across multiple range bins simultaneously. As seen in Fig. \ref{fig:multi}, when two users are more than 1 m apart, \name can monitor the HRR of both users simultaneously with negligible degradation in performance.

\subsubsection{Different Types of HR Variations} When measuring HRR, we find that the HR may increase or remain stable for a certain duration after exercise, rather than immediately commencing a decrease. An illustrative example of HR initially rising and then falling is depicted in Fig. \ref{fig:inc_hr}, it shows that the HR estimation obtained by \name maintains accuracy across various HR transitions, encompassing increases, stabilization, and decreases. We manually select the subset of data where an increase in HR is observed and calculate the mean HR estimation error of this dataset as 3.18 bpm. This value closely aligns with the overall estimation error.

%% file: chapter/5-relatedwork.tex
\section{Related Work\label{relatedwork}}


\subsection{Cardiac Activity Monitoring based on mmWave}

The mmWave sensing has gained significant attention in various sensing applications\cite{10193776, yuan2023detection, jia2023mmhawkeve}. Owing to the sub-millimeter motion resolution, mmWave signals have shown promise in measuring cardiac activity, which has been extensively studied in recent years.

Some state-of-the-art studies\cite{xu2021cardiacwave, zhang2022can} the mmWave signals to extract fine-grained characteristics of cardiac activity. For example, RF-SCG\cite{ha2020contactless} proposes a 4D Cardiac
Beamformer component to discover the 3D location of the heart. It further utilizes a CNN-based method to recover the seismocardiography (SCG) waveform from the reflected signals. On the other hand, some works\cite{zhao2020heart,wang2020vimo} focus on various practical application scenarios. mmECG\cite{xu2022mmecg} use modified VMD methods to extract the driver's vital signs from the complex driving environment. MoVi-Fi\cite{chen2021movi} achieves motion-robust vital signs waveform recovery through a deep contrastive learning approach. \cite{gong2021rf} combines direct sensing results from static instances and indirect prediction based on movement power estimation to achieve continuous vital sensing. 

In contrast, this paper focuses on monitoring HRR after exercise. This task presents significant challenges, including the respiratory harmonic interference and the non-stationary property of heartbeat signals. These key challenges are particularly pronounced in the considered scenario but are not easily addressed by existing works.

\subsection{Other Cardiac Activity Monitoring Methods} 

\textit{Contact cardiac activity monitoring.} The traditional methods for HRR estimation require users to wear electrodes to collect ECG after exercise\cite{fletcher2001exercise}. Such devices are usually expensive and necessitate medical professionals to operate. As an alternative, wearable sensors incorporating photoplethysmography (PPG) or ECG technology, such as smartwatches and chest straps, have gained popularity. Nonetheless, the requirement of wearing these devices during exercise imposes additional burdens on users.

\textit{Contactless cardiac activity monitoring.} The utilization of contact-based sensing techniques in practical applications is often constrained by various limitations, making contactless sensing a preferred option for better user experiences in many applications\cite{chen2018deepphys, liu2018monitoring, zheng2020v2ifi,yimiao2023aim, wang2017phasebeat, pai2021hrvcam,  weiguo2023micnest}. At present, contactless methods can generally be categorized into three main groups: vision-based, acoustics-based, and RF-based.

The vision-based methods rely on capturing periodic changes in the microscopic color of the subcutaneous surface to extract cardiac activity information such as HR\cite{chen2018deepphys} and HRV\cite{pai2021hrvcam}. Such methods are susceptible to light conditions and raise privacy concerns. Existing work on acoustic-based HR sensing \cite{wang2023df} utilizes ultrasonic signals to monitor subtle movements of the human body induced by heartbeat. Most of them suffer from noise interference and strong attenuation. The RF-based methods utilize various wireless signals, such as WiFi\cite{liu2018monitoring,wang2017phasebeat} and UWB\cite{zheng2020v2ifi}, to measure the human chest displacement for heartbeat estimation. Due to complex multipath, their performance may degrade as the environment changes.




%% file: chapter/6-conclusion.tex
\section{Conclusion\label{conclusion}}

In this paper, we propose \name, a contactless technique for monitoring HRR based on mmWave radar. To mitigate interference caused by respiratory harmonics, we introduce customized preprocessing techniques and an adaptive VMD method for extracting the heartbeat signal. Subsequently, we estimate the HR from the non-stationary heartbeat signal using a novel peak counting algorithm. Our experiment results show that \name achieves accurate and robust estimation of HR across diverse environmental conditions.